%% file: main-plain-style.tex
\documentclass[letterpaper, 11pt]{article}

\usepackage{geometry}
\usepackage{url}
\usepackage[numbers,square,sort,compress]{natbib}
\usepackage{graphicx}
\usepackage{amsmath}
\usepackage{amssymb}
\usepackage{algorithm}
\usepackage{enumitem}
\usepackage{booktabs}
\usepackage{hyperref}
\usepackage{tikz}
\usepackage{caption, subcaption}
\usepackage{algorithmicx,algpseudocode}
\usepackage{rotating}
\usepackage{pdflscape}
\usepackage{multirow}
\usepackage{tabularx}
\usepackage{changepage}

\long\def\comment#1{}

\title{A Survey on Fault-tolerance in \\ Distributed Optimization and Machine Learning
\thanks{This uploaded version fixes a citation \cite{liu2021asynchronous}.}}

\author{Shuo Liu \\ Department of Computer Science\\ Georgetown University \\ Washington DC, USA\\ \texttt{sl1539@georgetown.edu} }
\date{}

\providecommand{\iprod}[2]{\ensuremath{\left\langle #1,\,#2  \right\rangle}}
\providecommand{\norm}[1]{\ensuremath{\left\lVert#1\right\rVert }}
\providecommand{\mnorm}[1]{\ensuremath{\left\lvert#1\right\rvert}}
\providecommand{\dist}[2]{\ensuremath{\mathrm{dist}\left( #1,\,#2 \right)}}

\providecommand{\gradfil}[1]{\ensuremath{\textsf{GradFilter}\left( #1 \right)}}

\def\R{\mathbb{R}}
\def\Z{\mathbb{Z}}
\def\H{\mathcal{H}}
\def\B{\mathcal{B}}

\def\G{\mathcal{G}}

\def\expectation{\mathbb{E}}
\def\g{\boldsymbol{g}}

\def\V{\mathcal{V}}

\def\V{\mathcal{V}}
\def\E{\mathcal{E}}
\def\N{\mathcal{N}}
\def\A{\boldsymbol{A}}

\newtheorem{definition}{Definition}

\begin{document}

\maketitle    

\pagenumbering{roman}

\begin{abstract}
The robustness of distributed optimization is an emerging field of study, motivated by various applications of distributed optimization including distributed machine learning, distributed sensing, and swarm robotics. With the rapid expansion of the scale of distributed systems, resilient distributed algorithms for optimization are needed, in order to mitigate system failures, communication issues, or even malicious attacks. This survey investigates the current state of fault-tolerance research in distributed optimization, and aims to provide an overview of the existing studies on both fault-tolerant distributed optimization theories and applicable algorithms. 
\end{abstract}

\newpage
\tableofcontents
\newpage


\pagenumbering{arabic}  

\input{intro-arxiv}
\input{preliminaries-arxiv}
\input{byzantine-arxiv}
\input{others-arxiv}
\input{conclusion-arxiv}

\section*{Acknowledgements}
This survey is supported by a Fritz Fellowship from Georgetown University.

\bibliographystyle{plainnat}
\bibliography{bib}

\appendix
\input{appendix-arxiv}

\end{document}

%% file: intro-arxiv.tex
\section{Introduction}
\label{cpt:intro}

With rapid development in communication, sensing and learning systems, development in computation and storage capacity of computer systems, and growth in data collection, the problems in networked systems have gained significant attention \cite{dong2018theory, gao2019reinforcement, ren2010distributed, yuan2016bayesian, zhu2015distributed}. A \textit{networked system}\footnote{In this survey, without specification, the phrase \textit{networked system} is interchangeable to \textit{distributed system} and \textit{multi-agent system}, whilst ``network system'' stresses on the communication network, ``distributed system'' stresses on the distributed nature of the task, and ``multi-agent system'' stresses on the number of members participating in the system.} typically consists of a number of \textit{agents}, which work collaboratively to achieve certain global objective. Many problems in networked systems can be solved in the framework of optimization \cite{boyd2011convex}. The distributed nature of these problems and performance limitations of centralized strategies lead to increasing interest in distributed approaches to solve the optimization problems \cite{boyd2011convex, nedich2015convergence}. \\

Among all the topics in researches of distributed systems, the robustness of networked systems received recent research attention \cite{duchi2011dual, kailkhura2015consensus, marano2008distributed}. People have realized, in both theory and practice, that agent failures or adversarial behaviors of some agents may render non-robust distributed algorithms useless.
In the context of distributed optimization, the problem of \textit{fault-tolerance}, sometimes also \textit{resilience} or \textit{robustness}, becomes increasingly important. People wish to solve optimization problems distributed, while also being able to counter agent failures, communication issues, malicious attackers, etc., while utilizing computational power from various sources \cite{su2015byzantine1, su2015byzantine2, su2015fault3, su2015fault4, su2016fault, sundaram2015consensus}. This survey intends to study the current state of researches in fault-tolerant distributed optimization problems.

\subsection{Distributed Optimization}
\label{sec:dist-opt}

Generally, an optimization problem intends to find optima (maximum or minimum point(s)) of a given objective function \cite{floudas2008encyclopedia}. In the problem of distributed optimization, we consider a multi-agent system (or \textit{network}) of $n$ agents, and each agent $i$ has a local cost function $Q_i(x)$, where $x\in\R^d$ is the optimization variable, a $d$-dimensional vector of real values. The objective of distributed optimization is to minimize a global objective function, which is the aggregation of cost functions of all agents:
\begin{equation}
    \min_{x\in\R^d}\sum_{i=0}^nQ_i(x),
\end{equation}
while the algorithm is typically in a distributed manner, i.e., with local computation by the agents and communication among the agents. In practice, an algorithm of distributed optimization can also be required to output only one of the minimum points to the global objective function when there are multiple possible solutions, i.e., output a vector $x^*$ such that
\begin{equation}
    x^*\in\arg\min_{i\in\R^d}\sum_{i=0}^nQ_i(x).
    \label{eqn:obj-all-functions}
\end{equation}

The problem of optimization with multiple agents has been studied since the end of last century in the context of parallel and distributed computation \cite{bertsekas2003parallel, tsitsiklis1984problems, tsitsiklis1986distributed}. In recent years it has gained renewed interest due to its applications in various fields, including communication networks, power systems, sensor networks, and machine learning \cite{cao2012overview, nedich2015convergence, nedic2018improved, sayed2014adaptation, gao2019reinforcement}. Recent reviews of distributed optimization include following surveys and books: \cite{yang2019survey, boyd2011convex, nedich2015convergence, nedic2018distributed, sayed2014adaptation, giselsson2018large, ren2010distributed}.

\subsection{Fault-tolerance in distributed optimization}
\label{sec:intro-fault-tolerant}

The studies on fault-tolerance start from the increasing need of robustness in distributed systems. In many applications of distributed system,  the reliability of such a system can be affected by component failures, e.g., malfunctioning agents, communication difficulties, or malicious attacks, e.g., some agents intend to sabotage the collective effort by sending false information \cite{castro1999practical, kotla2007zyzzyva, brooks1996robust, blanchard2017machine}; therefore, 
the central goal regarding robustness of distributed systems is to make the automated systems able of making correct decisions in  presence of faulty data  \cite{brooks1996robust}. In the context of general distributed systems, the two major tools for solving the problem algorithmically are \textit{Byzantine agreement} \cite{lamport2019byzantine, dolev1982byzantine} and \textit{sensor fusion} \cite{marzullo1990tolerating, chew1991masking}. Byzantine agreement allows a distributed decision-making system tolerate a certain number of failed agents, while sensor fusion enhances the accuracy of sensors (or agents) by deploying sensors redundantly. \\

In the context of distributed optimization, fault-tolerance is also gaining more attention for the same reason. 
Correspondingly, there are also two major lines of work. One intends to allow the algorithm to produce correct output in presence of a certain number of failed agents. This includes countering faulty agents regardless of their behavior, i.e., \textit{Byzantine fault-tolerance}, or \textit{Byzantine resilience} \cite{su2016fault, sundaram2018distributed, gupta2020fault}, or modeling and countering certain kinds of failures or adversarial behaviors \cite{duchi2011dual, kailkhura2015consensus}. The other line of work intends to counter faulty agents by assigning redundant workloads to agents \cite{chen2018draco, rajput2019detox}. \\

There are different distributed optimization models considered by the fault-tolerance problem. Some analyses take the communication topology of the network into consideration \cite{sundaram2018distributed, zhao2019resilient, su2020byzantine}, while other researches focus on one or two major system architectures \cite{gupta2020fault, blanchard2017machine, chen2017distributed}. Some papers consider only the case in which the cost functions are scalar-valued \cite{su2016fault, sundaram2018distributed}. Some works are interested in the fault-tolerant problems in specific distributed optimization tasks (e.g., distributed machine learning), or specific distributed optimization methods (e.g., distributed gradient descent or distributed stochastic gradient descent) \cite{gupta2020byzantine, chen2017distributed, liu2021approximate}. \\

This survey intends to investigate the current state of researches in the fault-tolerant distributed optimization problem. The rest of this survey is organized as follows. In Section~\ref{chpt:preliminary} we revisit some related concepts and notations in graph theory and Byzantine consensus, the two related fields. In Section~\ref{chpt:byzantine} we discuss the existing research in Byzantine fault-tolerance problem of distributed optimization, including the problem formulation, solvability, and algorithms. We also discuss the special case of Byzantine fault-tolerant distributed machine learning. We then discuss some other fault-tolerance problems in distributed optimization, including different adversary models, and possible combination of privacy-preservation and fault-tolerance in Section~\ref{chpt:others}. Finally, we summarize this survey in Section~\ref{chpt:summary} and discuss possible future work.

%% file: preliminaries-arxiv.tex
\section{Preliminaries}
\label{chpt:preliminary}

In this section, we introduce some backgrounds related to fault-tolerance in distributed optimization.

\subsection{Graph theory}
\label{sec:graph-theory}
We revisit some basics in graph theory \cite{godsil2001algebraic} that is also used in modeling and analysis of fault-tolerant distributed optimization algorithms. Let $\G=(\V,\E)$ denote a \textit{directed graph} with the set of nodes (in our case, agents) $\V=\{1,...,n\}$ and the set of edges $\E\subseteq\V\times\V$. $(i,j)$ denotes a directed edge from node $i$ to node $j$. A graph becomes \textit{undirected} if and only if $(i,j)\in\E$ always implies $(j,i)\in\E$. For simplicity, we assume in this survey that no directed graph contains self loop, i.e., $(i,i)\notin\E$ for all $i\in\V$, unless specified otherwise. The \textit{in-neighbors} of node $i$ consists of all nodes that can transmit information to node $i$ directly, denoted by the set $\N_i^{\mathrm{in}}=\{j\in\V|(j,i)\in\E\}$. In contrast, the \textit{out-neighbors} of node $i$ consists of all nodes that $i$ can transmit information to directly, denoted by the set $\N_i^{\mathrm{out}}=\{j\in\V|(i,j)\in\E\}$. 
For undirected graphs, we denote $\N_i=\N_i^{\mathrm{in}}=\N_i^{\mathrm{out}}$ as the \textit{neighbors} of node $i$. \\

A directed \textit{path} from node $i_1$ to $i_k$ is a sequence of nodes $\{i_1,...,i_k\}$ such that $(i_j,i_{j+1})\in\E$ for $j=1,...,k-1$; and if such a path exists, we say $i_k$ is reachable from $i_1$. An undirected graph is \textit{connected} if for any $i,j\in\V$, $i\neq j$, there exists a path between $i$ and $j$. A directed graph is \textit{strongly connected} if every node is reachable from every other node. 
The vertex cut $S\subset\V$ of a connected graph of nodes that by removing the nodes in $S$ and edges connected to them, the graph becomes unconnected. The (vertex) \textit{connectivity} of a connected graph is the size of the graph's minimum vertex cut. A \textit{source component} in a directed graph is a subset of nodes, in which each node has a path to every other nodes in the graph. A \textit{connected dominating set} of an undirected connected graph is the set of nodes in which every node not in the set has a neighbor in the set, and the set of nodes with their edges forms a connected subgraph itself. \\

A communication network may also be modeled with time-varying feature, denoted by $\G(t)=(\V,\E(t))$, where the edge set can change over time. Path, connectivity, and neighborhood sets defined above can also be adapted to this model with the time stamp $t$. 

\subsection{Byzantine consensus}
\label{sec:byzantine-consensus}

Although Byzantine fault-tolerant distributed optimization is proposed quite recently \cite{su2015byzantine1}, the question of Byzantine consensus has been proposed and well studied for some time \cite{dolev1982byzantine, lamport2019byzantine, leblanc2013resilient, pasqualetti2011consensus, lynch1996distributed}. It would be useful for us to revisit the concepts of consensus and Byzantine consensus, the two basic problems in distributed computing. \\

In the original \textit{consensus} problem (sometimes \textit{agreement} problem) in a multi-agent system, each of the $n$ agents starts with an input of either 0 or 1. By communications between agents, the goal is for all the agents to eventually decide on a value in $\{0,1\}$ that satisfies the following conditions:
\begin{quote}
\begin{description}[nosep,leftmargin=*]
    \item[Agreement] No two agents decide on different values;
    \item[Validity] If all agent start with the same value, they decide on that value, and
    \item[Termination] All agents eventually decide.
\end{description}
\end{quote}
Intuitively, a consensus protocol allows a group of agents in a multi-agent system to agree on a single value, after certain amount of communications. \\

The \textit{Byzantine consensus} problem expands the original problem that instead of 0 or 1, each agent can now hold a value $v\in V$, where $V$ is a set of allowed values. Also, some agents in the system can be non-compliant to the prescribed algorithm and exhibit arbitrary behavior, including starting in an arbitrary state, sending arbitrary message, and making arbitrary updates to its value. These agents are called \textit{Byzantine faulty} agents. Under this setting, the goal is to satisfy the following correctness conditions:
\begin{quote}
\begin{description}[nosep]
    \item[Agreement] No two non-faulty agents decide on different values;
    \item[Validity] If all non-faulty agent start with the same value $v\in V$, they decide on the value $v$, and
    \item[Termination] All non-faulty agents eventually decide.
\end{description}
\end{quote}
The major change here comparing to plain consensus is that it is only reasonable to apply correctness conditions on all non-faulty agents in Byzantine consensus. Studies on Byzantine consensus include necessity and sufficiency conditions for solving the problem, and practical algorithms under different system architectures. Readers can refer to \cite{lynch1996distributed} for detailed introduction on Byzantine consensus. \\

The consensus problem can be further extended to consensus on multi-dimensional values (i.e., vectors) \cite{vaidya2013byzantine, vaidya2014iterative}, asymptotic or approximate consensus with reasonable definition of acceptable outcome \cite{tseng2014iterative, amelina2015approximate, leblanc2013resilient, fugger2018fast}, and their corresponding Byzantine version \cite{leblanc2013resilient, mendes2015multidimensional, vaidya2013byzantine}. Results in Byzantine consensus are important in research of Byzantine distributed optimization, since Byzantine optimization can in fact be viewed as a special case of Byzantine consensus \cite{su2015byzantine1}.


\subsection{Notations}
\label{sec:notations}

We summarize the notations frequently used in this survey in Table~\ref{tab:notations} for reference. The table includes both notations already introduced and those that will be introduced in the following sections.

\begin{table}[h]
    \centering
    \caption{Notations used in this survey.}
    \begin{tabular}{c|l}
        \toprule
        \textbf{Notation} & \textbf{Meaning} \\
        \midrule
        $\R$ & The set of real numbers. \\
        $\R^d$ & The set of $d$-dimensional real-valued vectors. \\
        $\R^{d\times n}$ & The set of real-valued matrices with $d$ rows and $n$ columns. \\
        $\Z$ & The set of integers. \\
        \midrule
        $\G(\V,\E)$ & The graph $\G$, with node (agent) set $\V$ and edge set $\E$. \\
        $(i,j)\in\E$ & A directed edge from node $i$ to node $j$ in $\E$. \\
        $\N_i^{\mathrm{in}},\,\N_i^{\mathrm{out}},\,\N_i$ & In-, out-neighbors, and neighbors of node $i$. \\
        $\G(t)=(\V,\E(t))$ & The graph $\G$ at time stamp $t$. \\
        \midrule
        $n$ & The number of agents in a distributed system. \\
        $f$ & The upper-bound of the number of faulty agents in a distri- \\
         &  buted system. \\
        $t$ & The iteration number in an iterative algorithm. \\
        \midrule
        $x[k]$ & The $k$-th element of a vector $x$. \\
        $Q_i(x)$ & The cost function of agent $i$. \\
        $v_i^t$ & Some vector $v$ produced by agent $i$ at time $t$. \\
        \midrule
        $\boldsymbol{A},\,\boldsymbol{A}_{ij}$ & A matrix $\boldsymbol{A}$ and the element of $\boldsymbol{A}$ at position $(i,j)$. \\
        \midrule
        $\B$ & The set of faulty (adversarial) agents. \\
        $\H$ & The set of non-faulty (honest) agents. \\
        $z\sim\mathcal{D}$ & A random variable $z$ drawn from distribution $\mathcal{D}$. \\
        \midrule 
        $\mnorm{\cdot}$ & Absolute value or cardinality of a set. \\
        $\norm{\cdot}$ & Some kind of norm. \\
        $\iprod{\cdot}{\cdot}$ & Inner product of two vectors. \\
        $\expectation$ & Expectation. \\
        $\dist{\cdot}{\cdot}$ & Hausdorff distance between two sets (including points). Also \\
         & see Appendix~\ref{appdx:definitions}. \\
        $\nabla f(x)$ & Derivative of a function $f(x)$. \\
        \bottomrule
    \end{tabular}
    \label{tab:notations}
\end{table}

%% file: byzantine-arxiv.tex
\section{Byzantine fault-tolerant distributed optimization }
\label{chpt:byzantine}

The majority of current studies on the fault-tolerant distributed optimization problem assumes the faulty agents to be \textit{Byzantine}, i.e., there is no assumption made on the behavior of faulty agents in the system. This problem is often referred to as the \textit{Byzantine fault-tolerant}, \textit{Byzantine-resilient}, or \textit{Byzantine-robust} distributed optimization problem. The advantage of formulating the problem this way is obvious: by making no behavioral assumption on faulty agents, the applicability of the results is quite broad. It is also a reasonable choice, since in practice, the trusted parts often cannot predict what faulty agents would do.

\subsection{Problem formulation}
\label{sec:formulation}

\citet{su2015byzantine1} formally proposed the Byzantine fault-tolerant distributed optimization problem of a sum of convex cost functions with real-valued scalar input and output in a complete communication graph. Recall in Section~\ref{sec:dist-opt} that we discussed the formulation of distributed optimization, where each agent $i$ in the system of $n$ agents has a cost function $Q_i(x)$, and the goal is to find a point $x^*$ such that 
\begin{equation}
    x^*\in\arg\min_{x}\sum_{i=0}^nQ_i(x). 
    \label{eqn:nft-goal}
\end{equation} In presence of faulty-agents, it is impractical to achieve the goal stated in \eqref{eqn:obj-all-functions}, since it is possible that the non-faulty agents can never know any information regarding cost functions of the faulty agents. \\

Suppose among the $n$ agents in the system, up to $f$ agents may be Byzantine faulty. Let $\V=\{1,...,n\}$ denote the set of all agents, $\B$ denote the set of faulty agents, and $\H=\V-\B$ be the set of non-faulty agents. The problem assumes $\mnorm{B}\leq f$. Also suppose $Q_i:\R\rightarrow\R$ for all $i$. \citet{su2015byzantine1} showed that the goal of finding a minimum point for the averaged cost functions of all non-faulty agents
\begin{equation}
    \widetilde{x}\in\arg\min_{x\in\R}\dfrac{1}{\mnorm{\H}}\sum_{i\in\H}Q_i(x). 
    \label{eqn:ft-goal-1}
\end{equation}
is also impossible. However, it is possible to achieve the following goal, 
\begin{align}
    &\widetilde{x}\in\arg\min_{x\in\R}\sum_{i\in\H}\alpha_iQ_i(x), \nonumber \\
    \textrm{such that }~&\forall i\in\H,~\alpha_i\geq0,~\textrm{and}~\sum_{i\in\H}\alpha_i=1.
    \label{eqn:ft-goal-2}
\end{align}
i.e., find a minimum point of a \textit{convex combination} of non-faulty agents' cost functions. Ideally, we want all $\alpha_i=\dfrac{1}{\mnorm{\H}}$, effectively goal \eqref{eqn:ft-goal-1}. Therefore, we want to maximize the number of $\alpha_i$'s bounded away from 0. The authors proved in \citep{su2015byzantine1} that such a goal can be achieved with certain restrictions when defining ``bounded away from 0'', and the maximum achievable number of bounded away weights (i.e., $\alpha_i$'s) is $\mnorm{\H}-f$. The results are further extended to arbitrary directed networks in \citep{su2015byzantine2}. \\

We can easily generalize the ideal goal \eqref{eqn:ft-goal-1} to real-valued multivariate cost functions: suppose for every agent $i$, its cost function $Q_i:\R^d\rightarrow\R$, the goal becomes
\begin{equation}
    \widetilde{x}\in\arg\min_{x\in\R^d}\dfrac{1}{\mnorm{\H}}\sum_{i\in\H}Q_i(x). 
    \label{eqn:ft-goal-1-mul-d}
\end{equation}
Comparing to the scalar version, this problem has wider applicability. However, due to the impossibility of its scalar counterpart, solving \eqref{eqn:ft-goal-1-mul-d} is not generally achievable either. 
Note that optimization goal \eqref{eqn:ft-goal-1-mul-d} on the \textit{averaged} cost functions is the same as the following
\begin{equation}
    \widetilde{x}\in\arg\min_{x\in\R^d}\sum_{i\in\H}Q_i(x),
    \label{eqn:ft-goal-1-mul-d-alt}
\end{equation}
i.e., the \textit{aggregated} cost functions, since a positive scaling coefficient does not affect the optimum. \\

For clarification of the concepts, we call the set of the minimum points 
\begin{equation}
    X^*=\arg\min_x\sum_{i\in\H}Q_i(x)
\end{equation}
the \textit{true} minimum point set. Correspondingly, the point of that set $x^*\in X^*$ is called a \textit{true minimum point}. For any subset of agents $S\subseteq\V$, we also use the following notations: 
\begin{equation}
    X_S=\arg\min_x\sum_{i\in S}Q_i(x)
\end{equation}
for the minimum point set of the aggregated cost functions of agents in $S$, and $x_s\in X_S$ for a minimum point in $X_S$. \\

Note that without specifying the property of cost functions of agents in an arbitrary subset $S$, the property of the minimum point set $X_S$ is also undecided, including but not limited to the following possible cases: (1) \mnorm{X_S}=0, i.e., there is no minimum point; (2) \mnorm{X_S}=1, i.e., there is only one minimum point; (3) the set $X_S$ forms a region or shape in the Euclidean space $\R^d$, or (4) the set $X_S$ consists of many unconnected parts. 

\subsection{Solvability: the importance of redundancy in cost-functions}
\label{sub:redundancy}

Since it is impossible to solve the Byzantine fault-tolerant distributed optimization problem \eqref{eqn:ft-goal-1-mul-d} in general, researchers eyed on reasonable extra conditions that can help solving the problem. A line of work shows the relationship between solvability and redundancy in cost functions. The cost functions of agents, although not identical, have underlying connections with each other. This kind of conditions can help achieving the optimization goal without requiring side knowledge of the agents. \\

\citet{su2015byzantine2} in their extension of their previous work \cite{su2015byzantine1} analyzed the scenario where the cost function $Q_i(x)$ of each agent $i$ is formed as a convex combination of $k$ \textit{input functions}. Formally, a function $h:\R\rightarrow\R$ is \textit{admissible} if $h(x)$ is convex, $L$-Lipschitz continuous, and its minimum point set $\arg\min h(x)$ is non-empty and compact. Given $k$ admissible input functions $h_1(x),...,h_k(x)$, there exists a matrix $\boldsymbol{A}\in\R^{k\times n}$, such that the cost function $Q_i(x)$ of each agent $i\in\V$ is of the form
\begin{equation}
    Q_i(x)=\A_{1i}h_1(x)+\A_{2i}h_2(x)+...+\A_{ki}h_k(x),
\end{equation}
where $\A_{ji}\geq0$ and $\sum_{j=1}^k\A_{ji}=1$ for all $i\in\V$ and $j=1,...,k$. The goal is to find a solution $\widetilde{x}\in\R$ to the following optimization problem
\begin{equation}
    \widetilde{x}\in\arg\min_{x\in\R}h(x)=\dfrac{1}{k}\sum_{j=1}^kh_j(x)
    \label{eqn:ft-goal-redundency-1}
\end{equation}
instead of \eqref{eqn:ft-goal-1}, in presence of up to $f$ Byzantine faulty agents. The authors considered both cases where the agents are aware or unaware of the matrix $\A$. Suppose agents are aware of $\A$, \eqref{eqn:ft-goal-redundency-1} can be achieved if $\A$ can correct up to $f$ arbitrary entry-wise errors as stated in \cite{candes2005decoding} and the communication graph admits \textit{Byzantine broadcast} \cite{bracha1987asynchronous}. On the other hand, suppose agents are not aware of $\A$ beforehand, \eqref{eqn:ft-goal-redundency-1} can also be solved if all input functions share at least one common minimum point, and there is a necessity condition of $n>3f$ setting an upper bound for this special case. \\

\citet{gupta2020fault,gupta2020resilience} studied a different type of redundancy in cost functions named $2f$-redundancy, defined as follows:
\begin{definition}[$2f$-redundancy]
    \label{def:2f-redundancy}
    The cost functions of a set of non-faulty agents $\H$ is said to satisfy $2f$-redundancy if for every subset $S\subseteq\H$ of size at least $n-2f$,
    \begin{equation}
        \arg\min_x\sum_{i\in S}Q_i(x)=\arg\min_x\sum_{i\in\H}Q_i(x).
    \end{equation}
\end{definition}
Note that here $x\in\R^d$ for any $d\in\Z_{>0}$. This redundancy condition implies that the aggregated cost functions of every $n-2f$ non-faulty agents minimizes at the same set of points. It is further showed that it is only possible to achieve goal \eqref{eqn:ft-goal-1-mul-d} in presence of up to $f$ Byzantine faulty agents if the non-faulty cost functions satisfy $2f$-redundancy. Note that this result does not require convexity of the cost functions. The authors also pointed out that although the conditions mentioned in Definition~\ref{def:2f-redundancy} appears technical, in many practical applications, such redundancy in cost functions ``occurs naturally'' \cite{gupta2020fault}, including applications in distributed sensing and distributed learning. \\

\citet{liu2021approximate} further generalized the redundancy notation to $(2f,\epsilon)$-redundancy, where $\epsilon$ is an approximate parameter, defined as follows:
\begin{definition}[\textbf{$(2f,\epsilon)$-redundancy}]
\label{def:approx_red}
The agents' cost functions are said to have {\em $(2f, \, \epsilon)$-redundancy} property if and only if for every pair of subsets $S, \,\widehat{S} \subseteq \{1, \ldots, \, n\}$ with
$\mnorm{S} = n-f$, $\mnorm{\widehat{S}} \geq n-2f$ and $\widehat{S} \subseteq S$,
\begin{equation}
    \dist{\arg\min_{x \in \R^d}\sum_{i\in S}Q_i(x)}{\arg\min_{x \in \R^d}\sum_{i\in \widehat{S}}Q_i(x)} \leq \epsilon. \label{eqn:red_dist}
\end{equation}
\end{definition}
where $\dist{\cdot}{\cdot}$ is Hausdorff distance\footnote{Defined in Appendix~\ref{appdx:definitions}} between two sets in $d$-dimensional Euclidean space. Intuitively, this redundancy condition implies that every set of no less than $n-2f$ non-faulty agents has its minimum point set of aggregated cost functions within $\epsilon$ distance to the true minimum point set. The authors showed that $(2f,\epsilon)$-redundancy is necessary and sufficient for a deterministic algorithm to output a point close enough (also measured by $\epsilon$) to a true minimum point. Similarly, this finding also does not require the cost functions to be convex, but only requires a compact minimum point set.

\subsection{Practical solutions to Byzantine fault-tolerant distributed optimization problems}
\label{sec:algorithms}

After analyzing the theoretical solvabitlity of Byzantine fault-tolerant distributed optimization problems, we further dive in to the algorithms that can be used to solve the problems in practice. 

\subsubsection{Commonly-studied system architectures}

Before we start, it is useful to discuss first the different system architectures those algorithms are designed for. As a recently started research topic, researchers make various assumptions to the system. But most of those assumptions can be summarized into the following categories.

\paragraph{Network architectures}

\begin{figure}[t]
    \centering
    \begin{subfigure}[b]{.45\textwidth}
    \includegraphics[width=\linewidth]{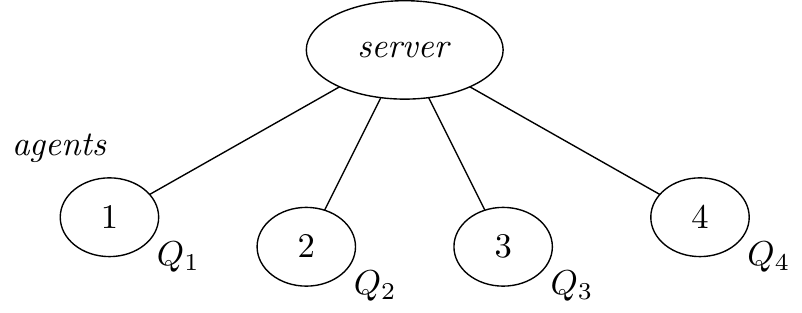}
    \caption{Server-based}
    \end{subfigure}
    \begin{subfigure}[b]{.45\textwidth}
    \includegraphics[width=\linewidth]{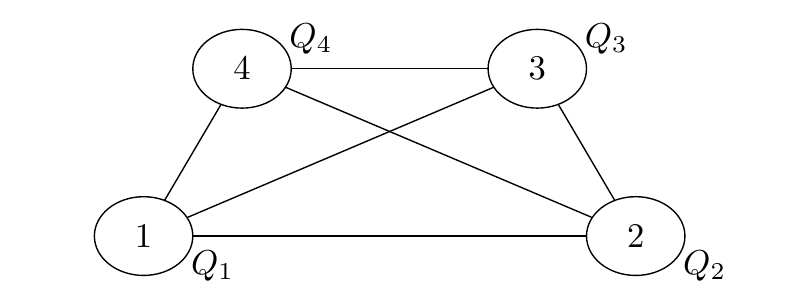}
    \caption{Peer-to-peer}
    \end{subfigure}
    \caption[Illustrations of the two major network architectures]{Illustrations of the two major network architectures in Byzantine fault-tolerant distributed optimization problems.}
    \label{fig:architecture}
\end{figure}

There are two major architectures studied in the researches of Byzantine fault-tolerant distributed optimization, illustrated in Figure~\ref{fig:architecture}: (1) the server-based architecture (or centralized architecture) and (2) the peer-to-peer architecture (or decentralized architecture). 
In the \textit{server-based} architecture, in the majority of researches, there is one server and $n$ agents; the agents communicate with the server, but not with each other. It is usually assumed that the server is trustworthy, but up to $f$ agents may be Byzantine faulty. There are also some works that use multiple servers or different communication models in order to achieve other specific goals. Unless specified otherwise, when referring to the server-based architecture in this survey, we are referring to the former common architecture.
In the \textit{peer-to-peer} architecture, $n$ agents are connected with each other, and up to $f$ of those agents may be Byzantine faulty. Note that the network is not necessarily complete, undirected, or fixed in the peer-to-peer architecture. \\

Problems under the two architectures can sometimes be equivalent. For example, provided that $f<\dfrac{n}{3}$, any algorithm for the server-based architecture can be simulated in a complete-graph peer-to-peer architecture using Byzantine broadcast primitive \cite{lynch1996distributed}. Therefore, some algorithms applicable to server-based architecture are also theoretically applicable to complete-graph peer-to-peer architecture.

\paragraph{Byzantine status}

We briefly mention that for the majority of the algorithms, the Byzantine status of agents may change during the execution of an algorithm, i.e, at different times, different agents may be Byzantine faulty, but the total number of agents exhibiting Byzantine behavior at any given time is bounded by $f$. An example in practice for the reason of this assumption is that any agent in a distributed system can experience system failure, but statistically, the total number of failed agents can be bounded by a certain number. Unless otherwise specified, the algorithms we introduced below do not require fixed Byzantine status. \\

However, in some researches, it is also useful to assume that Byzantine faulty agents are fixed during an execution. This kind of assumptions are necessary for some algorithms to achieve their fault-tolerance goal. We will discuss several algorithms of such kind in the following sections.

\paragraph{Data distributions}

In a general distributed optimization problem, each agent $i$ has its own cost function $Q_i(x)$ that is potentially different from other agents'. If in server-based network, the server does not have a cost function, nor does it have the knowledge of cost functions of agents. \\

In distributed learning setting, however, there might be different assumptions. The assumptions mainly fall in these categories: (1) all agents have their data samples drawn from the same data distribution $\mathcal{D}$; (2) each agent $i$ has their data drawn from a data distribution $\mathcal{D}_i$, or (3) all agents have the same dataset (parallel setting), including the cases in server-based network, the server also has the same dataset, or the server assigns (sends) data samples to agents during the training process. We will discuss fault-tolerance in distributed learning further in Section~\ref{sec:byzantine-learning}.

\subsubsection{Gradient descent with gradient filters}
\label{sub:gradient-filters}

One major category of these algorithms is called \textit{gradient filters} \cite{gupta2020byzantine}, or \textit{robust gradient aggregation} \cite{chen2017distributed, blanchard2017byzantine}, which are designed and used mainly with (distributed) gradient descent (abbr. DGD)  \cite{nedic2009distributed}. DGD is a popular method solving distributed optimization problems, hence it is natural to build upon it to achieve Byzantine fault-tolerance. \\

DGD is originally designed for peer-to-peer architecture, where each agent keeps its own local estimate $x_i$. During the algorithm, the agents communicate in iterations. In each iteration $t$, each agent $i$ shares its local estimate $x_i$ with other agents, then performs a consensus step and then a descent step. Specifically, the agent conducts the following update:
\begin{equation}
    x_i^{t+1}=\sum_{j=1}^nw_{ij}^tx_j^t-\eta_tg_i^t,
    \label{eqn:dgd-p2p-update}
\end{equation}
where $x_i^t\in\R^d$ indicates agent $i$'s local estimate at iteration $t$, $w_{ij}^t$ is the weight of communication link on edge $(j,i)\in\E(t)$ at iteration $t$ such that $\boldsymbol{W}(t)=\left(w_{ij}^t\right)\in\R^{n\times n}$ is doubly stochastic for all $t$, $g_i^t$ is the (sub)gradient\footnote{DGD is originally proposed for convex but not necessarily differentiable cost functions.} of the local cost function $Q_i(x)$ at $x_i(t)$, and $\eta_t>0$ is a diminishing step size\footnote{Defined in Appendix~\ref{appdx:definitions}}. For server-based architecture, DGD can also be adapted \cite{li2014scaling}. \\

One simple version of server-based DGD is described in Algorithm~\ref{alg:dgd-server}. Generally speaking, in each iteration $t$, each agent receives the current estimate $x^t$ from the server, computes and sends its update $g_i^t$ back to the server; the server keeps the current estimate $x^t$, computes its update in iteration $t$ using all update vector received from the agents, and updates its current estimate according to \eqref{eqn:honest-update}. \\

\begin{algorithm}[t]
	\caption{DGD for server-based architecture}
	\label{alg:dgd-server}
	\begin{algorithmic}[1]
	    \Procedure {DgdServer}{} \Comment{Server executes this procedure}
        \State \textbf{Initialize}: arbitrarily select an initial estimate $x^0$
        \State \textbf{Loop} until convergence. For current iteration $t=0,1,2,...$:
        \State \qquad Broadcast the current estimate $x^t$ to all agents
        \State \qquad Wait until receive gradient $g_i^t$ from each agent $i$ \label{line:dgd-receive}
        \State \qquad Update the current estimate $x^t$ by summing the updates of all agents:
            \begin{equation}
                x^{t+1}=x^t-\eta_t\sum_{i=1}^tg_i^t
                \label{eqn:honest-update}
            \end{equation}
        \EndProcedure
        \Statex
        \Procedure{DgdAgent}{} \Comment{Each agent $i$ executes this procedure}
            \State \textbf{Loop} until server stops. For each iteration $t$:
            \State \qquad Wait until receive current server estimate $x^t$
            \State \qquad Compute (sub)gradient of local function $Q_i(x)$ at $x^t$: \label{line:dgd-agent-compute}
                \begin{equation}
                    g_i^t= \nabla Q_i(x^t)
                    \label{eqn:honest-agent}
                \end{equation}
            \State \qquad Send $g_i^t$ as agent $i$'s update to the server
        \EndProcedure
	\end{algorithmic}
\end{algorithm}

Byzantine agents in the system may not follow the prescribed algorithms. Specifically, in line~\ref{line:dgd-agent-compute} of Algorithm~\ref{alg:dgd-server}, instead of computing and sending its gradient following \eqref{eqn:honest-agent}, a Byzantine agent may choose to send arbitrary information as its update. When the server receives $g_i^t$'s from agents in line~\ref{line:dgd-receive}, some of them may be incorrect. In fact, \citet{blanchard2017machine} showed that no linear combination methods for aggregating the updates (including summing and averaging, indicated by \eqref{eqn:honest-update}) can tolerate even a single Byzantine agent. Thus comes the \textit{gradient filters}, methods that redesign the aggregation rule. A general framework for server-based Byzantine fault-tolerant optimization with DGD, or Byzantine gradient descent (BGD), is shown in Algorithm~\ref{alg:bgd-server}. \\

\begin{algorithm}[t]
	\caption{BGD framework for server-based architecture}
	\label{alg:bgd-server}
	\begin{algorithmic}[1]
	    \Procedure {BgdServer}{} \Comment{Server executes this procedure}
        \State \textbf{Initialize}: arbitrarily select an initial estimate $x^0$
        \State \textbf{Loop} until convergence. For current iteration $t=0,1,2,...$:
        \State \qquad Broadcast the current estimate $x^t$ to all agents
        \State \qquad Wait until receive gradient $g_i^t$ from each agent $i$ \label{line:dgd-receive}
        \State \qquad Update the current estimate $x^t$ by summing the updates of all agents:
            \begin{equation}
                x^{t+1}=x^t-\eta_t\,\gradfil{g_1^t,...,g_n^t}
                \label{eqn:ft-update}
            \end{equation}
        \EndProcedure
        \Statex
        \Procedure{BgdAgent}{} \Comment{Each \textit{non-faulty} agent $i$ executes this procedure}
            \State \textbf{Loop} until server stops. For each iteration $t$:
            \State \qquad Wait until receive current server estimate $x^t$
            \State \qquad Compute (sub)gradient of local function $Q_i(x)$ at $x^t$: \label{line:dgd-agent-compute}
                \begin{equation}
                    g_i^t= \nabla Q_i(x^t)
                    \label{eqn:ft-agent}
                \end{equation}
            \State \qquad Send $g_i^t$ as agent $i$'s update to the server
        \EndProcedure
	\end{algorithmic}
\end{algorithm}

The major difference between Algorithms~\ref{alg:bgd-server} and DGD is \eqref{eqn:ft-update}: instead of updating the estimates by the sum of all agents updates, BGD uses an aggregation rule $\gradfil{\cdot}$. Generally speaking, $\textsf{GradFilter}:\R^d\times n\rightarrow\R^d$ is a function that takes $n$ vectors of $d$-dimension, and output a vector of $d$-dimension. Normally, the $n$ vectors would be those updates received from $n$ agents at iteration $t$, i.e., $\left\{g_i^t\right\}_{i=1}^n$, and the output is used as the server's update at iteration $t$. Also note that only non-faulty agents follow the procedure $\textsc{BgdAgent}$ and send their correct gradient as its update. Byzantine agents may behave arbitrarily, including but not limited to sending its correct update, sending an arbitrary $d$-dimensional vector, sending other arbitrary information, or not sending anything. Note that only an arbitrary $d$-dimensional vector would confuse the server, since other behaviors will immediately indicate the agent deviating from the prescribed algorithm, and therefore must be faulty; if the Byzantine status of agents is fixed, the algorithm can further remove the ill-behaved agent in future iterations. \\

Suppose out of $n$ agents, up to $f$ can be Byzantine faulty. We introduce some gradient filters as follows. For simplicity, we denote $\g^t=\left\{g_i^t\right\}$ as the set of update vectors unless otherwise specified. 

\paragraph{Krum} Krum \cite{blanchard2017machine} is an angle-based (or direction-based) gradient filter, built upon a Krum score $s(i)$. Suppose among the set of vectors $\g^t$, $i\rightarrow j$ denotes the fact that $g_j^t$ belongs to the $n-f-2$ closest vectors to $g_i^t$, with distance between two vectors defined by some norm $\norm{\cdot}$. The score $s(i)=\sum_{i\rightarrow j}\norm{g_i^t-g_j^t}^2$, i.e., the sum of the distances between $g_i^t$ and the $n-f-2$ vectors closest to it. Then, 
\begin{equation}
    \textsf{GradFilter}\left(g_1^t,\dots,g_n^t\right)=g_{i_*}^t,    
\end{equation}
where $i_*$ is an agent with the smallest score $s(i_*)$. Krum outputs one of the vectors submitted by the agents. The expected time complexity of computing this filter is $O(n^2d)$. Krum is originally proposed as a distributed machine learning filter for stochastic gradient descent (SGD), but it can also be applied to general optimization problems.

\paragraph{Multi-KRUM} Multi-Krum \cite{blanchard2017machine, blanchard2017byzantine} is a variant of Krum. Instead of selecting one vector, multi-Krum selects $m$ vectors and averages them, where $m$ is a hyperparameter. There are two versions, the first version (also known as $m$-Krum) iterates $m$ times, each time it calculates the scores for all vectors in $\boldsymbol{g}^t$, selects a vector with the highest score and remove it from the set $\boldsymbol{g}^t$; the second version selects $m$ vectors with the $m$ smallest scores. Obviously, the time complexity of the second version is the same as Krum, which is significantly better than that of the first version. 

\paragraph{Coordinate-wise methods} Coordinate-wise median and trimmed mean \cite{yin2018byzantine} process the received vectors by coordinates. Suppose $v[k]$ denotes the $k$-th coordinate of a vector $v$, the gradient filter can be written as follows
\begin{align}
    \begin{split}
        &\gradfil{g_1^t,...,g_n^t}=v,~\textrm{where} \\
        &v[k]=\left\{\begin{array}{cl}
            \textsf{median}\left\{g_1^t[k],...,g_n^t[k]\right\}, & \textrm{for median}, \\
            \textsf{trmean}_\beta\left\{g_1^t[k],...,g_n^t[k]\right\}, & \textrm{for trimmed mean}, \\
        \end{array}\right.
    \end{split}
\end{align}
where $\textsf{median}\{\cdot\}$ finds the median from a set of scalars, while $\textsf{trmean}_\beta\{\cdot\}$ first drop the smallest and largest $\beta$ fraction, then computes the mean of the rest of the values from a set of scalars. $\beta$ is a hyperparameter, which is required to be larger than the fraction of faulty agents $\alpha$, i.e., $\beta\geq\alpha=f/n$. Note that coordinate-wise median does not require a known fraction of faulty agents. \\

Phocas \cite{xie2018phocas} is a variant of trimmed mean method. The authors also noted that for coordinate-wise methods, the number of faults can also be viewed coordinate-wisely, i.e., the update vectors from every agent can be corrupted, so long as for each coordinate $k\in\{1,...,d\}$, the number of faulty values in the set $\{g_i^t[k]\}_{i=1}^n$ is upper-bounded by $f$.
Another coordinate-wise method, mean around median \cite{xie2018generalized}, calculates each coordinate $k$ by the mean of $n-f$ values $g_i^t[k]$'s that are closest to the median of all values. These filters are also originally proposed as a distributed learning filter for SGD. 

\paragraph{Geometric median} Geometric median also has some robustness \cite{rousseeuw1985multivariate, mhamdi2018hidden, chen2017distributed, xie2018generalized}, and sometimes is used as a benchmark comparing other gradient filters. Geometric median can be computed to arbitrary precision $\epsilon$ in nearly linear time $O(nd\log^3\dfrac{1}{\epsilon})$ \cite{cohen2016geometric}. That being said, in practice the (geometric) median-based aggregation still dominates the training time in large-scale settings \cite{chen2018draco}. 

\paragraph{Median of means} Geometric median of means \cite{chen2017distributed} extends the statistical estimator to vectors. Specifically, suppose $b$ divides $n$, agents are divided into $k=n/b$ groups, each of size $b$. In each iteration, the gradient filter finds the mean vector in each group, and then uses the geometric median of those $k$ mean vectors as its update. Formally,
\begin{equation}
    \gradfil{g_1^t,...,g_n^t}=\textsf{median}\left\{\frac{1}{b}\sum_{j=1}^bg_j^t,...,\frac{1}{b}\sum_{j=(k-1)b+1}^ng_j^t\right\}.
\end{equation}
The number of groups $k$ is a hyperparameter, which should be chosen such that $k>2f$. This filter is originally proposed as a distributed learning filter for SGD.

\paragraph{MDA} Minimum-diameter averaging \cite{el2020genuinely}, also called \textit{Brute} \cite{mhamdi2018hidden}, is another median-based method, originally proposed as a statistical estimator in \cite{rousseeuw1985multivariate}. It uses the average of the subset $S$ of $n-f$ vectors which has the minimum diameter. Formally,
\begin{equation}
    \begin{split}
        &\gradfil{g_1^t,...,g_n^t}=\dfrac{1}{n-f}\sum_{i\in S}g_i^t,~\textrm{where} \\
        &S=\arg\min_{\begin{subarray}{c}T\subset\V\\\mnorm{T}=n-f\end{subarray}}\left(\max_{i,j\in T}\norm{g_i^t-g_j^t}\right).
    \end{split}
\end{equation}
The computation complexity of MDA is $O\left({n \choose f} + n^2d\right)$. This filter is originally proposed as a distributed learning filter for SGD.

\paragraph{Norm-based methods} Norm filtering (or comparative gradient elimination, CGE) and norm-cap filtering (or comparative gradient clipping, CGC) \cite{gupta2020resilience, gupta2019byzantine, gupta2020byzantine} are a group of filters looking at the norms of the vectors. Intuitively, these filters keep the $n-f$ vectors in $\g^t$ with smallest norms, while dropping the rest vectors (elimination) or scaling the rest so that they also have the  $n-f$-largest norm (clipping). Formally, suppose the vectors in $\g^t$ are sorted as follows:
\begin{equation}
    \norm{g_{i_1}^t}\leq...\leq\norm{g_{i_{n-f}}^t}\leq\norm{g_{i_{n-f+1}}^t}\leq...\leq\norm{g_n^t},
\end{equation}
the gradient filters can be written as follows:
\begin{equation}
    \gradfil{g_1^t,...,g_n^t}=\left\{\begin{array}{cl}
        \displaystyle\sum\limits_{j=1}^{n-f}g_{i_j}^t, & ~\textrm{for CGE}, \\
        \displaystyle\sum_{j=1}^{n-f}g_{i_j}^t + \sum_{j=n-f+1}^n\dfrac{\norm{g_{i_{n-f}}^t}}{\norm{g_{i_j}}}g_{i_j}^t, & ~\textrm{for CGC}.
    \end{array}\right.
\end{equation}
The computational complexity of the two methods are both $O(n(\log{n}+d))$. Intuitively, even if a gradient from faulty agent is kept, its norm is bounded by a gradient from non-faulty agent, and therefore could not do much damage to the server's update. The two methods are originally proposed in the context of distributed linear regression, and analyzed also in general distributed optimization \cite{gupta2020fault} and distributed learning \cite{gupta2020byzantine}.

\paragraph{Bulyan} Bulyan \cite{mhamdi2018hidden} is a meta-aggregation rule that can be applied on another gradient filters $\textsf{GradFilter}$. Bulyan consists of two steps. In the first step, Bulyan iterates $n-2f$ times; in each iteration, it runs $\textsf{GradFilter}$ on the set $\g^t$, selects a $g_i^t$ closest to $\textsf{GradFilter}$'s output, add it to a set $S$, and remove it from $\g^t$. After the first step, the set $S$ contains $n-2f$ vectors. In the second step, Bulyan generates a new $d$-dimensional vector coordinate-wisely from the $n-2f$ vectors with a median-based method, and uses it as the update. It is designed such that it has provable resilience property even if $\textsf{GradFilter}$ in use does not have such a property (e.g., geometric median). \\

It is worth noting that the convergence analyses of most of the gradient filters requires standard assumptions like continuity, differentiability, Lipschitz smoothness, convexity, strong convexity, including those that are proposed for distributed learning or SGD. However, the most popular machine learning methods, including deep neural networks, are known to be generally non-convex \cite{collobert2006trading, jain2017non}. \\

It is also worth noting that some gradient filters such as coordinate-wise trimmed mean and CGE are also analyzed under the peer-to-peer setting \cite{gupta2021byzantine}. \\

For reference purpose, we summarize all gradient filters listed in this section in Table~\ref{tab:grad-filters} in Appendix~\ref{appdx:tables}. 

\subsubsection{Gradient coding}
\label{sub:coding}

One of the applications of coding theory is to enable error correction \cite{hamming1986coding}. Gradient coding is originally proposed as a straggler mitigation method \cite{tandon2017gradient}, which is used to speed up synchronous distributed first-order methods \cite{cotter2011better, raviv2020gradient, charles2017approximate}. Several works build upon it and extend it to the adversarial setup. Generally, this type of methods work under the parallelization setting of distributed learning, where agents (and server, if any) all have the same dataset.

\paragraph{Draco} Draco \cite{chen2018draco} is a gradient-coding-based method, in which the server assigns same data samples to (on average) $r$ agents, agents compute multiple stochastic gradients and send the gradients to the server in a coded way. The server receives the coded messages from the agents, and decodes the message to recover the correct gradients and identifies the faulty messages. With proper-selected coding method, Draco can tolerate up to $(r-1)/2$ Byzantine agents with linear-time encoding and decoding. The assignment of same tasks to multiple agents is also known as \textit{algorithmic redundancy}. DETOX \cite{rajput2019detox} is an extension of Draco that combines algorithmic redundancy with robust aggregation, with increased speed and improved robustness. 

\paragraph{Randomized reactive redundancy} After Draco and DETOX, \citet{gupta2019randomized} proposed a similar coding-based framework, in which instead of applying coding every iteration, the server invokes the coding scheme (i.e., check for faults) with probability $q>0$, while in other iterations, the server simply carries out DGD. By properly choosing the value of $q$, this framework reduces the computational overhead caused by coding to arbitrarily small. The scheme is especially effective when Byzantine agents are fixed, since once detected, the faulty agents will be removed. The authors also proposed heuristic checking by server and combination with gradient filters.

\subsubsection{Other methods}
\label{sub:other-methods}

There are some other methods that do not fall into the above major categories. Though those methods are more or less related to previous methods. 
\paragraph{Zeno} Zeno \cite{xie2018zeno} is a gradient aggregation rule based on the fact that the server obtains certain number of data samples, designed for distributed learning with SGD. In each iteration, instead of assigning data samples to agents, the server calculates a \textit{reliability} score of the vectors sent from the agents, based on the fact that all the data samples are drawn from the same data distribution $\mathcal{D}$. The server then aggregate using those $n-f$ most reliable gradients to obtain its update.

\paragraph{One-round robust aggregation} Instead of checking or filtering every iteration, \citet{yin2018byzantine} propose a Robust One-round Algorithm, in which the non-faulty agents conduct their own optimization process using their local cost functions separately, and send their final estimates to the server; Byzantine agents may send arbitrary information. The server then aggregates the estimates (including faulty ones) to compute its own final estimate. There is no communication between agents and the server the whole process except the final step. In \cite{yin2018byzantine} geometric median is used as the final aggregation rule, and convergence analysis is provided. The method is proposed originally for distributed learning, and suppose all agents have the same data distribution, the method can achieve comparable empirical performance to some gradient filters. 

\paragraph{Variance reducing techniques} There are certain techniques used in machine learning known to have the ability of reducing the variance of stochastic gradients. Since the stochastic gradients are generated from randomly drawn data samples, a variance-reduce method is possible to speed up the training process by ``stablizing'' the gradients \cite{cutkosky2019momentum, qian1999momentum}. \citet{gupta2020byzantine} empirically studied that averaging historical gradients and increasing batch size helps reducing the variance of stochastic gradients, and therefore boosts the performance of fault-tolerant gradient filters. \citet{karimireddy2020learning} showed that using momentum \cite{qian1999momentum} helps achieving provable convergence of any Byzantine robust gradient filter, while \citet{el2020distributed} studied the boosting of robustness by momentum when computed at agents.

\subsubsection{Studies under peer-to-peer architecture}

Many methods, including the most of aforementioned gradient filters are proposed under the server-based system architecture. Although we already mentioned in the beginning of this section that we can simulate server-based algorithms under peer-to-peer architecture using Byzantine broadcast primitive, it is worth noting that there is also research specifically studies the peer-to-peer, or decentralized system architecture. 
Note that in peer-to-peer settings the agents do not broadcast gradients, but rather their local estimates to other agents (recall \eqref{eqn:dgd-p2p-update}, peer-to-peer DGD). \\

\citet{su2020byzantine} showed that, based-on Byzantine consensus results on peer-to-peer networks \cite{vaidya2012matrix}, a distributed optimization problem with scalar local cost functions has a Byzantine-resilient gradient descent algorithm if the network has a non-empty source component after removing all faulty agents and their edges. \citet{sundaram2018distributed} proposed \textit{Local Filtering (LF) Dynamics}, a protocol that allows Byzantine fault-tolerant distributed optimization in a $f$-local network, i.e., each non-faulty agent $i$ has up to $f$ faulty neighbors in $\N_i$, if the network satisfies certain connectivity property called $(r,s)$-robustness. \\

\citet{gupta2020byzantine_decentralized} studied the $2f$-redundancy property in decentralized system. Specifically, the authors proposed \textit{Comparative Elimination} (abbr. CE) method, similar to gradient filters under server-based architecture, as a decentralized optimization algorithm for a fully-connected network. The method will mitigate the detrimental impact of potentially incorrect values from Byzantine faulty agents.

\subsection{Byzantine fault-tolerant distributed learning}
\label{sec:byzantine-learning}

Distributed machine learning is a popular subproblem of distributed optimization. Naturally, a lot of work in Byzantine fault-tolerant distributed optimization also focuses on the case of distributed learning. The formulation of Byzantine fault-tolerant distributed learning problem is related to that of distributed optimization, but with its own characteristics. \\

One typical formulation of the problem can be described as follows \cite{yin2018byzantine, gupta2020byzantine, gupta2020fault, mhamdi2018hidden, chen2017distributed}. Suppose there are $n$ agents in the system, out of them up to $f$ can be Byzantine faulty. The training data samples are drawn i.i.d. from some unknown data distribution $\mathcal{D}$. A machine learning model $\Theta$ has a learning parameter $x$ in the form of $d$-dimensional vectors. The model $\Theta$ decides a loss function $\ell(x;z)$ for each data point $z\sim\mathcal{D}$. Let $\expectation_{z\sim\mathcal{D}}$ denote the expectation with respect to the random data sample $z$. The goal is to minimize with respect to $x$ the function 
\begin{equation}
    Q(x)\triangleq\expectation_{z\sim\mathcal{D}}\left[\ell(x;z)\right],
\end{equation}
namely the population cost (or loss) function, i.e. find a point $\widehat{x}$ such that
\begin{equation}
    \widehat{x}\in\arg\min_{x\in\R^d}Q(x).
\end{equation}

Comparing this formulation with the general fault-tolerant distributed optimization goal \eqref{eqn:ft-goal-1-mul-d}, we see that here, $Q_i(x)=Q(x)$ for every agent $i$, therefore, $2f$-redundancy holds trivially. However, none of the agents, including the server, knows the cost function exactly. For gradient-based algorithms, DGD should be changed to distributed stochastic gradient descent (abbr. D-SGD) \cite{gupta2020byzantine}, where instead of gradient $g_i^t=\nabla Q_i(x^t)$, the agent $i$ draws one or some data samples $z$ or $\boldsymbol{z}$ and computes its stochastic gradient 
\begin{equation}
    g_i^t=\left\{\begin{array}{cl}
        \displaystyle\nabla\ell(x;z), & \textrm{when drawing one data sample, or} \\
        \displaystyle\nabla\ell(x;\boldsymbol{z})=\sum_{z\in\boldsymbol{z}}\ell(x;z), & \textrm{when drawing multiple data sample,}
    \end{array}\right.
\end{equation}
in each iteration $t$. \\

Another more general formulation \cite{liu2021approximate, he2020byzantine, pillutla2019robust, li2019rsa} assumes each agent $i$ has a potentially different data distribution $\mathcal{D}_i$. Then each agent $i$ has a local cost function
\begin{equation}
    Q_i(x)\triangleq\expectation_{z\sim\mathcal{D}_i}\left[\ell(x;z)\right].
\end{equation}
The goal is to find a point $\widehat{x}$ that minimizes the aggregated cost functions of non-faulty agents,
\begin{equation*}
    \widehat{x}\in\arg\min_{x\in\R^d}\sum_{i\in\H}Q(x).
\end{equation*}
Note this goal is the same as \eqref{eqn:ft-goal-1-mul-d-alt}. One application of this formulation is federated learning \cite{mcmahan2017communication}, where different agents have different data samples and underlying data distributions. Gradient filters like RSA \cite{li2019rsa}, RFA \cite{pillutla2019robust}, and RGE \cite{data2020byzantine} are proposed specifically under this formulation.

\subsection{Resilience notations}
\label{sec:resilience}

Given the facts that the original goal of distributed optimization \eqref{eqn:nft-goal} is not achievable in presence of Byzantine faulty agents, and that there are a variety of ways to measure the resilience of a Byzantine fault-tolerant distributed optimization algorithm, here we introduce some of them.

\paragraph{$(f,\epsilon)$-resilience} $(f,\epsilon)$-resilience \cite{liu2021approximate} is a resilience notation for Byzantine fault-tolerant distributed optimization algorithms. A deterministic algorithm is said to be $(f,\epsilon)$-resilient for some $\epsilon\geq0$, if the output of the algorithm $\widehat{x}$ satisfies the following:
\begin{equation}
    \dist{\widehat{x}}{\arg\min_{x\in\R^d}\sum_{i\in\H}Q_i(x)}\leq\epsilon.
\end{equation}
Intuitively, to be $(f,\epsilon)$-resilient, the output of the algorithm should be within $\epsilon$ distance to the true minimum point set. Furthermore, DGD with gradient filters, i.e., Algorithm~\ref{alg:bgd-server} can achieve $(f,\epsilon)$-resilience if the gradient satisfies certain conditions, with several continuity and convexity assumptions; those gradient filters include coordinate-wise trimmed mean and CGE \cite{liu2021approximate}. $(f,0)$-resilience is also called \textit{exact fault-tolerance} \cite{gupta2020fault}, requiring that the algorithm's output satisfies the goal \eqref{eqn:ft-goal-1-mul-d-alt} exactly. It is also shown that $(2f,\epsilon)$-redundancy and $2f$-redundancy are the necessary conditions to $(f,\epsilon)$-resilience and $(f,0)$-resilience, respectively \cite{liu2021approximate, gupta2020fault}.

\paragraph{$(\alpha,f)$-resilience} $(\alpha,f)$-resilience \cite{blanchard2017machine} is a notation used for measuring a gradient aggregation rule, under the same distribution setting of fault-tolerant distributed learning. Suppose a group of vectors $V_1,...,V_n\in\R^d$ are drawn i.i.d. from some distribution $G$, with $\expectation[G]=g$. Also suppose $B_1,...,B_f\in\R^d$ be a group of arbitrary vectors. An aggregation rule $\mathsf{GradFilter}$ is said to be $(\alpha,f)$-Byzantine resilient for some $0\leq\alpha<\pi/2$ if, for any $1\leq j_1\leq...\leq j_f\leq n$, the output vector of the aggregation rule
\begin{equation*}
    V=\textsf{GradFilter}(V_1,...,\underbrace{B_1}_{j_1},...,\underbrace{B_f}_{j_f},...,V_n)
\end{equation*}
satisfies the following:
\begin{enumerate}[nosep,label=(\roman*)]
    \item $\iprod{\expectation[V]}{g}\geq(1-\sin\alpha)\cdot\norm{g}^2>0$;
    \item For $r=2,3,4$, $\expectation\norm{V}^2$ is bounded above by a linear combination of terms $\expectation\norm{G}^{r_1}$,...,$\expectation\norm{G}^{r_{n-1}}$ with $r_1+...+r_{n-1}=r$.
\end{enumerate}

Intuitively, we want the output of a gradient filter to be close enough to the expected gradient, and the filter can control the effects of the discrete nature of SGD dynamics \cite{bottou1998online}. Some aforementioned gradient filters and others are known to be $(\alpha,f)$-Byzantine resilient, including Krum \cite{blanchard2017byzantine}, geometric median \cite{chen2017distributed, xie2018generalized}, coordinate-wise median \cite{xie2018generalized}, mean near median \cite{xie2018generalized}, and Bulyan \cite{mhamdi2018hidden}. 

\paragraph{$(\delta_{\max},c)$-robust aggregator} $(\delta_{\max},c)$-robust aggregator \cite{karimireddy2020learning} is another notation of measuring aggregation rules. Similarly, consider a group of independent random vectors $V_1,...,V_n$, such that a non-faulty subset $N\subseteq\{1,...,n\}$ of size $\mnorm{N}\geq(1-\delta)n$ satisfies that for any apriori fixed $i,j\in N$,
\(    \expectation[V_i]=\expectation[V_j]\), and \(\expectation\norm{V_i-V_j}^2\leq\rho^2 \)
for some $\rho$. An aggregation rule $\mathsf{GradFilter}$ is said to be $(\delta_{\max},c)$-robust if its output vector
\begin{equation*}
    V=\gradfil{V_1,...,V_n}
\end{equation*}
satisfies that for some constant $c$, $\expectation\norm{V-\dfrac{1}{\mnorm{N}}\sum_{i\in N}V_i}^2\leq c\delta\rho^2$. The authors then argued that combining $(\delta_{\max},c)$-robust aggregator and momentum SGD, an algorithm can solve distributed learning problems with non-convex smooth cost functions.

%% file: others-arxiv.tex
\section{Other fault-tolerant distributed optimization problems}
\label{chpt:others}

Outside the scope of our discussion in Byzantine fault-tolerance distributed optimization in Section~\ref{chpt:byzantine}, some other problems related to fault-tolerance in distributed optimization are also studied. 

\subsection{Alternative adversarial models}
\label{sec:alt-models}

There is a handful of recent researches considering specific adversarial models other than Byzantine models, certain types of attacks against distributed optimization algorithms, or analyzing possible behavior of an adversarial agent in the system. Such research are rather ad hoc or unstructured, but still provide important view points other than the common Byzantine model. \\

We use the similar notations as we used for Byzantine fault-tolerant distributed optimization problems in Section~\ref{chpt:byzantine}; specifically, out of all agents $\V$, $\H$ stands for the set of honest agents, and $\B$ stands for the set of adversarial agents, with $\mnorm{\V}=n$ and $\mnorm{B}\leq f$. \\

\citet{yin2019defending} considers \textit{saddle point attack} against existing Byzantine fault-tolerant distributed (non-convex) learning algorithms. Many machine learning models have non-convex cost functions. Although gradient descent or its variants are known to converge to a local minimum point with high probability \cite{jin2017escape, lee2016gradient}, Byzantine agents can manipulate those methods into a fake local minimum near a saddle point, i.e., saddle point attack. This is a specific type of attack that only happens when the cost function contains saddle points, at which the gradient of the cost function would also be 0, and therefore satisfies the stopping criteria of many Byzantine distributed learning algorithms. The authors proposed ByzantinePGD with \textit{perturbation} \cite{jin2017escape} to escape saddle point during the training process. \\

\citet{wu2018data} discussed a \textit{data injection attack} against distributed optimization with peer-to-peer DGD update \eqref{eqn:dgd-p2p-update}. Specifically, suppose the adversarial agents' goal is to steer the final estimate of all agents to a target $x\in\R^d$, an adversarial agent $i\in\B$ will not send its local estimate $x_i^t$ to other agents, but rather $x$ with artificial noise $z_i^t$, i.e. $x+z_i^t$. The adversarial agents will also try to behave as if they are converging, having $\lim_{t\rightarrow\infty}\norm{z_i^t}\overset{a.s.}{=}0$. The authors then proposed a local metric that trustworthy agents can compute to \textit{detect} (notice the existence of an adversary in its neighborhood) and \textit{localize} (distinguish which neighbor is adversarial) the adversarial agents performing such kind of attack. There is a group of research focusing on the detection of adversarial agents in distributed optimization \cite{ravi2019detection, li2020detect, wu2021detection}. \\

\citet{prasad2018robust} studied two specific types of statistical models: (1) arbitrary outliers in Huber's $\epsilon$-contamination model \cite{huber1965robust}, and (2) heavy-tails, i.e., the data distribution $\mathcal{D}$ has weak moment assumptions. The authors argued that such kind of statistical models are common in real-world datasets. The authors then introduced a class of robust estimators (gradient filters) with robustness guarantees for a variety of statistical models: linear regression, logistic regression, and exponential family models. Such robust estimators can be easily applied to distributed settings. \\

\citet{ravi2019case} analyzed possible behavior of malicious agents in the system. Suppose the malicious agents intend to manipulate its objective function, such that the output using cost functions from all agents $x^a$ will deviate from a correct output $x^*$ by a vector $\epsilon$, i.e. $x^a=x^*+\epsilon$. The authors derived that the magnitude of $\epsilon$ is bounded by a function of the number of faults $f$, and the gradients of the malicious agents. Therefore, in order to launch a substantial attack, either the value of $f$ needs to be large, or the gradients from malicious agents need to be large, which might become giveaways of malicious agents. \\

\citet{charikar2017learning} views the fault-tolerance learning problem from a data perspective. It is assumed that the $\alpha$ fraction of the data is drown from an unknown data distribution $\mathcal{D}$ and the rest $(1-\alpha)$ is not. The \textit{list-decodable} learning returns a list of $\textrm{poly}(1/\alpha)$ answers and one of them is correct. The \textit{semi-verified} learning suggests that if a small trusted dataset, also drawn from $\mathcal{D}$, is provided, it is possible to use the trusted dataset to enable accurate extraction of information from the larger dataset with untrusted data. Such findings can be applied to Byzantine fault-tolerant distributed learning problems. \\

Based on resilient consensus with trusted agents \cite{abbas2014resilient}, research shows that \textit{trusted agents} in the system can be crucial against adversarial agents \cite{zhao2019resilient, baras2019trust, emiola2021distributed}. \citet{baras2019trust} presented a trust-aware consensus algorithm in peer-to-peer networks that can effectively detect Byzantine adversaries and exclude them, even in sparse networks with connectivity less than $2f+1$. \citet{zhao2019resilient} presented another algorithm that if trusted agents induce a connected dominating set, the algorithm outputs a point bounded by the convex minimum point set of weighted average of all non-faulty agents' cost functions. 

\subsection{Fault-tolerance and privacy}
\label{sec:privacy}

Privacy issue in optimization has gained increasing attention in recent years, in both non-distributed settings \cite{abadi2016deep, damaskinos2020differentially, shoukry2016privacy, he2019privacy, song2013stochastic} and distributed settings \cite{naseri2020toward, tang2019privacy, lu2018privacy, xie2017privacy, jayaram2020mystiko}. In distributed settings, some research proposes encryption-based methods to prevent passive attackers from intercepting the exchanged information between agents in the network \cite{tang2019privacy, lu2018privacy}, while others utilizes \textit{differential privacy} \cite{naseri2020toward, xie2017privacy}, a gold standard notion for privacy-preserving in data \cite{dwork2014algorithmic}. Informally, a differential-private algorithm is insensitive to small differences in its input dataset. \\

Some recent research tries to simultaneously achieve privacy-preservation and fault-tolerance. \citet{he2020secure} presented a Byzantine-resilient and privacy-preserving machine learning solution with a two-server protocol. Achieving \textit{local differential privacy} \cite{evfimievski2003limiting}, the data of each agent is secure against any other agents in the system, and the two honest-but-curious servers. The Byzantine resilience can be provided by any gradient filter, e.g., those we mentioned in Section~\ref{sub:gradient-filters}, and the protocol achieves the same result as non-private algorithms using the same gradient filter. The authors also showed that their protocol only has negligible computation and communication overhead comparing to non-private methods. \\

\citet{so2020byzantine} proposed Byzantine-resilient secure aggregation (BREA) framework to achieve both privacy-preservation and fault-tolerance in federated learning. Different from the previous work, BREA only has one server in the system. The secrecy among agents is achieved by a verifiable secret sharing protocol \cite{feldman1987practical} ensuring that updates from an agent cannot be learnt by other agents, while fault-tolerant aggregation is managed by the server using a gradient filter such as Krum. It is worth mentioning though, BREA does not have a provable differentially-private property. \\

\citet{guerraoui2021differential} analyzed the compatibility between differentially-private noisy injection methods \cite{naseri2020toward, shokri2015privacy} and Byzantine-resilient gradient filters for distributed learning under a one-server architecture (i.e., the server-based architecture in Figure~\ref{fig:architecture}). The authors showed that in order to simultaneously guarantee Byzantine-resilience and differential privacy, the agents must sample data with batch size of the order of $\sqrt{d}$, where $d$ is the parameter size of the machine learning model. Such large batch size is often impractically large, since many state-of-the-art machine learning models have a huge number of parameters \cite{zagoruyko2016wide}. The authors further showed that, in strongly-convex cost function machine learning tasks, with differentially-private noise injection, the training error rate of a Byzantine-resilient gradient filter is of the order of $\dfrac{d}{b^2}$, where $b$ is the batch size; while non-private training error rate of the same gradient filters is independent from $d$. \\

Needless to say, the last results of compatibility from \cite{guerraoui2021differential} are rather frustrating. However, since this is an emerging research area, there are still many unexplored methods and mechanisms to be studied. As the authors of \cite{guerraoui2021differential} pointed out, Byzantine fault-tolerant methods other than gradient filters, and variance reduction techniques \cite{bottou2018optimization} both still show potential based on their analysis.

%% file: conclusion-arxiv.tex
\section{Summary}
\label{chpt:summary}

This survey summarizes the current state of studies in the fault-tolerance problem of distributed optimization, including both Byzantine fault-tolerant distributed optimization and other fault-tolerant distributed optimization researches. For Byzantine fault-tolerance distributed optimization, current researches studied the formulation and solvability of the problem; the practical solutions to the problem, including gradient filters, gradient coding methods, and other methods; the special case of Byzantine fault-tolerance in distributed learning; and commonly seen resilience notations characterizing Byzantine fault-tolerant distributed optimization algorithms. For other fault-tolerant distributed optimization researches, there is a group of work that proposed and studied some specific adversarial models. There is also an emerging line of work that intends to combine both robustness and privacy to distributed optimization algorithms. 

\subsection{Future work}

Based on the findings in our survey, there are various open questions in this field. We list some possible future work as follows.

\paragraph{{Gradient filters}} Although there already are a variety of gradient filters presented in this survey, many of them are either computationally extensive, or with weaker convergence property or stochastic error rate (when applying to D-SGD). It is still interesting to see if there are other gradient filters that can achieve both efficiency and correctness. One interesting idea, similar to Bulyan that applies the same filter multiple times, would be to see both theoretically and empirically effects of applying multiple different gradient filters in the same fault-tolerant algorithm, i.e, the effects of combinations. Also, efficiency could be achieved by heuristic filtering \cite{gupta2019byzantine}, instead of applying the gradient filter in every iteration.

\paragraph{{Peer-to-peer network}} The majority of Byzantine fault-tolerant algorithms are built under the server-based architecture, e.g., \cite{blanchard2017machine, gupta2020byzantine, chen2018draco}, etc.; while studies under peer-to-peer architectures, although exist \cite{gupta2020byzantine_decentralized, gupta2021byzantine}, are rather rare, and the results are not as systematic. It would be interesting to explore further how the communication network structure is related to solvability of the fault-tolerance problem, and also practical algorithms to achieve fault-tolerance. Recall the difference of DGD under peer-to-peer and server-based architectures, the most gradient filters and other methods cannot be directly applied to peer-to-peer settings.

\paragraph{{Asynchrony}} Asynchronous distributed optimization is a major branch of distributed optimization studies \cite{iutzeler2013asynchronous, srivastava2011distributed, zhang2014asynchronous}. Although some previous work suggested that their results can be easily extended to asynchronous setting, the combination of asynchrony and fault-tolerance is still a topic to be explored. It would be interesting to see some directed results on asynchronous systems, e.g., the effect of achieving both goals on issues such as convergence property. 
For example, \textit{$(f,r;\epsilon)$-redundancy}, an extension of $(2f,\epsilon)$-redundancy discussed in Section~\ref{sub:redundancy} is proposed in \citep{liu2021asynchronous}, which can be utilized to tackle both up to $r$ stragglers and up to $f$ Byzantine faulty agents at the same time.

\paragraph{{Privacy-preservation}} We already discussed the current attempts on combination of privacy-preservation and fault-tolerance. Still, this is an emerging topic with real-world applications and impact. 

\paragraph{{Adversary models}} The majority of the fault-tolerant optimization researches focus on Byzantine fault-tolerance, both in theory and in practice. However, in many real-world scenarios, such assumption may be too strong. A group of omnipotent faulty agents that have knowledge on the algorithm, status of other agents, or even all the data is unlikely in many cases. Instead, it is more likely that only a number of faulty agents can collaborate with each other, or can be corrupted by an adversary \cite{delporte2011disagreement}, or their adversarial behavior is limited. 

%% file: appendix-arxiv.tex
\section{Appendix: Definition of some mathematical concepts}\label{appdx:definitions}

In this appendix, we note some definitions of the mathematical concepts mentioned without explanation in the survey for readers' reference.

\subsection{Hausdorff distance}

To begin with, the \textit{distance} of two points $x,y$ in $d$-dimensional Euclidean space $\R^d$ induced by a given norm $\norm{\cdot}$ is
\begin{equation}
    \dist{x}{y}=\norm{x-y}.
\end{equation}
The distance between a point $x\in\R^d$ and a set $Y\subset\R^d$ can then be defined as
\begin{equation}
    \dist{x}{Y}=\inf_{y\in Y}\norm{x-y}.
\end{equation}
The Hausdorff distance between two set $X,Y\subset\R^d$ is then defined as follows:
\begin{equation}
    \dist{X}{Y}=\max\left\{\sup_{x\in X} \dist{x}{Y}, ~ \sup_{y\in Y}\dist{y}{X}\right\}. \label{eqn:hausdorff}
\end{equation}
Note that the definition of \ref{eqn:hausdorff} also applies to the distance between two points, or between a point and a set, if viewing a point as a set with one element.

\subsection{Diminishing step size}

A diminishing step size $\eta_t$ in the learning process is an sequence satisfying the following conditions \cite{bottou1998online}:
\begin{equation}
    \sum_{t=1}^\infty\eta_t=\infty,~\textrm{and}~\sum_{t=1}^\infty\eta_t^2<\infty.
\end{equation}

\section{Appendix: Summary table}
\label{appdx:tables}

\begin{landscape}
\thispagestyle{empty}
\begin{table}[p]
    \centering
    \fontsize{10pt}{11pt}\selectfont
    \caption[Summary of gradient filters]{Summary of gradient filters. Cells with ``-'' indicates that information is not provided in the original resource or other research to the best of our knowledge.}
    \label{tab:grad-filters}
    \begin{adjustwidth}{-1.4cm}{}
    \begin{tabularx}{1.05\linewidth}{c|ccccccc}
        \toprule
        \textbf{Gradient filter} & \textbf{Type} & \begin{tabular}{c}\textbf{Outputs an}\\\textbf{input vector}\end{tabular} & \begin{tabular}{c}\textbf{Time complexity}\\\textbf{at agents}\\\textbf{per iteration}\end{tabular} & \begin{tabular}{c}\textbf{Is}\\\textbf{$(\alpha,f)$-resilient}\end{tabular} & \begin{tabular}{c}\textbf{Originally}\\\textbf{designed for}\end{tabular}
        & \begin{tabular}{c}\textbf{Fault-tolerance}\\\textbf{threshold}\end{tabular}
        & \textbf{Comments} \\
        \midrule
         Krum & \multirow{3}{*}{\begin{tabular}{c}Angle-\\based\end{tabular}} & Yes & $O(n^2d)$ & Yes & D-SGD & \multirow{3}{*}{$f<(n-2)/2$} & \\
         $m$-krum & & No & $O(n^2d)$ & Yes & D-SGD & \\
         Multi-Krum & & No & $O(n^2d)$ & Yes & D-SGD & \\
        \midrule
         \begin{tabular}{c}Coordinate-wise\\median\end{tabular} & \multirow{5}{*}{\begin{tabular}{c}Coordinate-\\wise\end{tabular}} & No & $O(nd)$ & Yes & D-SGD & See \cite{yin2018byzantine} \\
         \begin{tabular}{c}Coordinate-wise\\trimmed mean\end{tabular} & & No & $O(nd)$ & Yes & D-SGD & $f<n/2$ \\
         Phocas & & No & $O(nd)$ & - & D-SGD & $f<n/2$ \\
         Mean around median & & No & $O(nd)$ & Yes & D-SGD & $f<n/2$ \\
        \midrule
         Geometric median & \multirow{6}{*}{\begin{tabular}{c}Median-\\based\end{tabular}} & No & $O\left(nd\log^3\dfrac{1}{\epsilon}\right)$ & - & - & - & \multirow{2}{*}{\begin{tabular}{l}$\epsilon$ as approx.\\param.\end{tabular}} \\
         Median of means & & No & $O\left(nd+fd\log^3\dfrac{1}{\epsilon}\right)$ & - & D-SGD & $f<n/2$ \\
         MDA & & No & $O\left({n \choose f} + n^2d\right)$ & Yes & D-SGD & $f\leq(n-1)/2$ \\
        \midrule
         CGC & \multirow{2}{*}{\begin{tabular}{c}Norm-\\based\end{tabular}} & No & $O\left((n+f)d+n\log{n}\right)$ & - & linear regression & $f<n/2$ \\
         CGE & & No & $O\left(n(\log{n}+d)\right)$ & - & DGD & $f<n/2$ \\
        \midrule
         Bulyan & \begin{tabular}{c}Meta-\\method\end{tabular} & No & $O((n-2f)C+nd)$ & Yes & D-SGD & $f\leq(n-3)/4$ & \begin{tabular}{l}$C$ is the com-\\plexity of a\\filter\end{tabular} \\
        \bottomrule
    \end{tabularx}
    \end{adjustwidth}
\end{table}
\end{landscape}